\documentclass[graybox]{svmult}

\usepackage{mathptmx}       
\usepackage{helvet}         
\usepackage{courier}        
\usepackage{type1cm}        
\usepackage{makeidx}         
\usepackage{graphicx,epsfig,color}
\usepackage{multicol}        
\usepackage[bottom]{footmisc}

\usepackage{amssymb}
\usepackage{amsmath}
\usepackage{enumerate}

\makeindex             

\begin{document}

\title*{Min-Max Regret Scheduling To Minimize the Total Weight of Late Jobs With Interval Uncertainty}
\titlerunning{Min-Max Regret Scheduling With Interval Uncertainty}
\author{Maciej Drwal}
\institute{Maciej Drwal \at Department of Computer Science, Wroclaw University of Science and Technology,\\ Wybrzeze Wyspianskiego 27, 53-370 Wroclaw, Poland\\ \email{maciej.drwal@pwr.edu.pl}}
%
%
\maketitle

\abstract{
We study the single machine scheduling problem with the objective to minimize the total weight of late jobs. It is assumed that the processing times of jobs are not exactly known at the time when a complete schedule must be dispatched. Instead, only interval bounds for these parameters are given. In contrast to the stochastic optimization approach, we consider the problem of finding a robust schedule, which minimizes the maximum regret of a solution. Heuristic algorithm based on mixed-integer linear programming is presented and examined through computational experiments.}


\section{Introduction}\label{sec:1}

We consider the following fundamental scheduling problem. A set of jobs is given to be processed on a single machine. Each job requires possibly different processing time to complete and cannot be interrupted or preempted. There is a fixed due-date until which the work should be finished. However, it is uncertain how much processing each of the task would exactly take. Before the schedule is dispatched on the machine, the only available data is the set of interval bounds, to which the actual processing requirements belong. The goal is to sequence the jobs, so that the number of the jobs that complete before the due-date is maximal (or, equivalently, the number of late jobs is minimal). In a more general problem variant, each job is associated with a weight (or cost), and the objective is to minimize the sum of weights of late jobs.

This problem arises in a many diverse application areas. For instance, this situation is experienced by a client who leases a fixed machine time (e.g., in a computing center) to carry out a number of tasks, but each of them requires unknown processing time; on the other hand, upper bounds on the processing times are set. This problem may also occur in a manufacturing process, when a fixed due-date is set for a batch of finished items to be delivered, but production time of each item may vary within known bounds.

The processing times uncertainty can be handled in a several different ways. One common approach is to use stochastic framework, and model the quantities of interest as random variables. This has its advantages in specific situations; however, it often brings the need for collecting data in order to estimate parameters. Moreover, in certain critical applications, the probabilistic guarantees, offered by such an approach, may not be sufficient. In this paper, we consider a {\it robust optimization} approach \cite{ben2009robust}, \cite{goerigk2016algorithm}. Each realization of uncertain parameters is treated as equally possible. Our aim is to come up with such a solution that degrades the least as compared to the best solution in every possible scenario. This measure of solution quality is reflected in the notion of {\it maximum regret} \cite{milnor1951games}.

Most of the basic scheduling problems have been already considered within the robust optimization framework \cite{kasperskiminmax}, \cite{aissi2009min}, \cite{drwal2016complexity}. The majority of these works concerns the more restrictive case of discrete uncertainty (a finitely many ways of parameter realizations). If the processing times were known precisely, the unweighted variant of the problem considered in this paper could be solved in polynomial time \cite{brucker2007scheduling}. However, even for 2 processing times scenarios, it becomes NP-hard \cite{aloulou2008complexity}. The case of interval processing times, described in the next section, appears to occur more naturally in practice. Although the number of processing times scenarios in such case is potentially infinite, solution algorithms may take the advantage of the structural information of uncertainty sets. Unfortunately, the problem with interval data is also NP-hard \cite{drwal2017b}, even if all weights are equal. Moreover, deterministic variant with arbitrary weights is already NP-hard. A viable solution approach is the application of mathematical programming techniques, presented in this paper.

\section{Problem Formulation}

The deterministic version of the considered scheduling problem is denoted $1|d_i=d|\sum w_iU_i$. Given is the set of jobs $J = \{ 1, 2, \ldots, n \}$. Each job $j \in J$ is described by the processing time $p_j$ and weight $w_j$. Let $d > 0$ denote the due-date. A solution (schedule) is a permutation $\pi = (\pi(1), \pi(2), \ldots, \pi(n))$, where $\pi(k)$ is the index of job scheduled to be executed as $k$th from the start. Equivalently, we encode the solution as a binary matrix ${\bf x}$, where $x_{kj} = 1$, iff $j$th job is scheduled on position $k$ from the start, and $x_{kj} = 0$ otherwise. The completion time of job scheduled on position $k$ is defined as:
$$
C({\bf x}, k) = \sum_{i=1}^k \sum_{j \in J} x_{ij} p_j.
$$
We define $U({\bf x}, k) = 0$, iff $C({\bf x}, k) \leq d$; we say that the job on position $k$ is {\it on-time}. Otherwise, $U({\bf x}, k) = \sum_{j \in J} x_{kj} w_j$, and we say that the job on position $k$ is {\it late}. An optimal schedule is one that minimizes the weighted number of late jobs, $F({\bf x}) = \sum_{k=1}^n U({\bf x}, k)$. An important special case, when $w_j=1$ for all $j \in J$, is the problem of minimizing only the number of late jobs.

In an uncertain problem, for each $j \in J$, instead of exact processing times $p_j$, we are given interval bounds $p^-_j, p^+_j$, so that the actual processing time can be any real number between them.
A vector of processing times will be called a {\em scenario}. The set of all possible scenarios is defined as:
$$
	\mathcal{U} = \{ {\bf p}=(p_1, \ldots, p_n) : \; \forall_{j \in J} \;\; p^-_j \leq p_j \leq p^+_j \}.
$$
The value of objective function in a scenario ${\bf p} \in \mathcal{U}$ will be denoted by $F({\bf x}, {\bf p})$.

Let $\mathcal{P}$ be the set of all $n$-by-$n$ permutation matrices. Given a solution ${\bf x} \in \mathcal{P}$, and a scenario ${\bf p} \in \mathcal{U}$, we define the regret as:
$$
	R({\bf x}, {\bf p}) = F({\bf x}, {\bf p}) - \min_{{\bf y} \in \mathcal{P}} F({\bf y}, {\bf p}).
$$
A schedule represented by matrix ${\bf y}$ in this context will be called an {\it adversarial} schedule. Then the maximum regret is denoted as:
\begin{equation}\label{max-regret}
	Z({\bf x}) = \max_{{\bf p} \in \mathcal{U}} R({\bf x}, {\bf p}).
\end{equation}
We will also use the notation $Z(\pi)$ to denote the maximum regret $Z({\bf x})$ of a matrix ${\bf x}$ equivalent to permutation ${\pi}$.

A scenario that maximizes the regret will be called a {\em worst-case scenario}.
A {\em robust optimal} solution ${\bf x}^*$ is one that minimizes the maximum regret:
\begin{equation}\label{min-max-regret}
Z^* = Z({\bf x}^*) = \min_{{\bf x} \in \mathcal{P}} Z({\bf x}).
\end{equation}

\section{Computation of Maximum Regret}

An essential prerequisite for solving the robust problem \eqref{min-max-regret} is the solution for the subproblem of regret maximization \eqref{max-regret}. Let us fix a schedule $\pi$. Since a due-date $d$ is common for all jobs, there exists a job on such a position $l$ in $\pi$, so that all jobs $\pi(1), \pi(2), \ldots, \pi(l-1)$, are on-time, while all jobs $\pi(l), \pi(l+1), \ldots, \pi(n)$, are late. Observe that worst-case scenario for $\pi$ is one for which the difference between the total weight of late jobs in $\pi$, and the total weight of late jobs in adversarial schedule is maximal. In the special case of equal weights, each late job contributes equally to the value of objective function, thus for any fixed scenario, an adversarial schedule is constructed by sorting all jobs with respect to nondecreasing processing times. This is not true for the case of general weights, where computing adversarial schedule for a fixed scenario is equivalent to solving an instance of knapsack problem.

Intuitively, in the worst-case schedules, the jobs that complete before the due-date $d$ would have the processing time closer to their respective upper bounds of uncertainty intervals. On the other hand, late jobs would generally have shorter worst-case processing times, closer to their lower bounds of uncertainty intervals. Such processing times allow for the late jobs to be early in the adversarial schedule, maximizing the number of on-time jobs. 

Let us consider the following example problem instance with $n=3$ identical jobs. Each has the same processing time interval $[p_j^-, p_j^+] = [1, 3]$, for $j \in \{ 1,2,3 \}$. Let the due-date be equal to 5.
Since the jobs are identical, the maximum regret is the same for each schedule, thus let $\pi = (1,2,3)$. It can be seen  that the following processing times constitute a worst-case scenario: $p_1 = 3$, $p_2 = 2+a$, for $a \in (0,1]$, and $p_3 = 1$. Only the first job completes on-time in schedule $\pi$. However, in an adversarial schedule $\pi'=(2,3,1)$, jobs 2 and 3 complete on-time, while only job 1 is late, giving the regret value 1. As shown in the example, for a given solution there may be infinitely many worst-case scenarios.

For any fixed schedule $\pi$ we can write a mixed-integer linear program (MIP), which allows to compute the worst-case processing times, as well as the value of maximum regret. The program is the following:


\begin{equation}\label{mip-1:obj}
	\textrm{maximize } \;\;\; \sum_{j \in J} w_j \left( z_j - q_j \right),
\end{equation}
subject to:
\begin{align}
	& \;\;\;  \sum_{j \in J} v_j \leq d, \label{mip-1:1}
\\
	\forall_{k=1,\ldots,n} & \;\;\; \sum_{i=1}^k p_{\pi(i)} + d_{\epsilon} q_{\pi(k)} \geq d_\epsilon, \label{mip-1:2}
\\
	\forall_{j \in J} & \;\;\; v_j - p_j^+ z_j \leq 0, \label{mip-1:3}
\\
	\forall_{j \in J} & \;\;\; p_j + p_j^+ z_j - v_j \leq p_j^+, \label{mip-1:4}
\\
	\forall_{j \in J} & \;\;\; v_j - p_j \leq 0, \label{mip-1:5}
\\
	\forall_{j \in J} & \;\;\; p_j^- \leq p_j \leq p_j^+,
\\
	\forall_{j \in J} & \;\;\; z_j \in \{ 0, 1 \}, q_j \in \{ 0, 1 \}. \label{mip-1:7}
\end{align}

Binary decision variable $z_j$ assumes value 1 if and only if job $j$ is on-time in an adversarial schedule is the worst-case scenario, and binary decision variable $q_j$ assumes value 1 if and only if job $j$ is on-time in $\pi$ in the worst-case scenario. Decision variable $p_j$ represents the worst-case processing time of $j$th job. Values of these variables are determined due to the set of constraints \eqref{mip-1:2}. These constraints are satisfied when $q_{\pi(k)} = 0$, for such $k$ that are on-time in $\pi$ in the worst-case scenario, and for $q_{\pi(k)}=1$ for such $k$ that are late. Constant $d_\epsilon = d + \epsilon$ in \eqref{mip-1:2}, where $\epsilon$ is a small positive value. Continuous variables $v_j$ are introduced to linearize the mixed terms $v_j = p_j z_j$, through the set of constraints \eqref{mip-1:3}--\eqref{mip-1:5}, as required for the constraint \eqref{mip-1:1} to be linear.

Note that although standard solution algorithms for this program may require time increasing exponentially in $n$, in practice it can be solved very quickly. Computational experiments indicate, for example, that for $n = 100$ optimal solutions can be computed in about one second on a modern computer, while even for thousands of jobs optimal solutions can be found within few minutes.

\section{Finding Robust Solutions}

We present a heuristic method that allows to determine solutions with low maximum regret for the problem \eqref{min-max-regret}. The method consists of two phases. In the first phase we try to determine a good initial solution, and in the second phase we use randomized local search in order to improve the initial solution.

The first phase is accomplished by solving a mixed-integer linear program that approximates the value of optimal robust solution. Let us consider a fixed schedule given by a permutation matrix ${\bf x}$. Since the optimization direction for robust schedule is the minimization, as opposed to the subproblem of maximization of regret \eqref{max-regret}, we form a dual program of the linear programming relaxation of \eqref{mip-1:obj}--\eqref{mip-1:7}. After relaxing \eqref{mip-1:7} to $0 \leq z_j \leq 1$ and $0 \leq q_j \leq 1$, for all $j \in J$, we can write:
\begin{equation}
	\textrm{minimize } \sum_{j \in J} \left( -d_\epsilon \lambda_j^{a} + p_j^+ \lambda_j^{b} - p_j^- \lambda_j^c + \lambda_j^d + \lambda_j^e + p_j^+ \lambda_j^h \right)+ d \lambda_0 \label{mip-2:obj}
\end{equation}
subject to:
\begin{align}
	\forall_{j \in J} & \;\;\; -\sum_{k=1}^n \sum_{i=1}^k x_{ij}\lambda_k^a + \lambda_j^b - \lambda_j^c -\lambda_j^g + \lambda_j^h \geq 0, \label{mip-2:1}
\\
	\forall_{j \in J} & \;\;\; -d_\epsilon \sum_{k=1}^n x_{kj} \lambda_k^a + \lambda_j^d \geq -w_j, \label{mip-2:2}
\\
	\forall_{j \in J} & \;\;\; \lambda_j^e - p_j^+ \lambda_j^f + p_j^+ \lambda_j^h \geq w_j, \label{mip-2:3}
\\
	\forall_{j \in J} & \;\;\; \lambda_j^f + \lambda_j^g - \lambda_j^h + \lambda_0 \geq 0. \label{mip-2:4}
\end{align}
Dual variable $\lambda_0$ corresponds to the constraint \eqref{mip-1:1}, while the subsequent sets of dual variables $\boldsymbol{\lambda}^a$ through $\boldsymbol{\lambda}^h$ correspond to the constraints \eqref{mip-1:2}--\eqref{mip-1:7}.

Since this is minimization program, we can also treat the matrix ${\bf x}$ as a decision variable, and solve this program for unknown ${\bf x}$, along with $\boldsymbol{\lambda}$, after adding the matching constraints:
\begin{align}
	\forall_{j \in J} & \;\;\; \sum_{i=1}^n x_{ij} = 1, \label{mip-2:5}
\\
	\forall_{i=1,\ldots,n} & \;\;\; \sum_{j \in J} x_{ij} = 1, \label{mip-2:6}
\\
	& \;\;\; x_{ij} \in \{ 0 , 1 \}. \label{mip-2:7}
\end{align}
Observe that in this case constraints \eqref{mip-2:1} and \eqref{mip-2:2} contain products of decision variables $x_{ij}$ and $\lambda_k^a$. However, since $x_{ij}$ are binary, and $\lambda_k^a$ are nonnegative continuous, we can linearize these products in a standard way, by substituting new variables $u_{kij} = \lambda_k^a x_{ij}$, and adding three sets of constraints, similar to \eqref{mip-1:3}--\eqref{mip-1:5}.

An optimal solution ${\bf x}$ of \eqref{mip-2:obj}--\eqref{mip-2:7} corresponds to an adversarial solution with fractional values of $z_j$ and $q_j$ (these are dual variables corresponding to \eqref{mip-2:2}--\eqref{mip-2:3}). In result, we get an upper bound on the optimal solution. This solution can be sometimes easily improved by rounding $z_j$ and $q_j$ to 0-1 values, and determining the corresponding ${\bf x}$ that satisfies \eqref{mip-1:1}--\eqref{mip-1:7}. We use the resulting binary matrix ${\bf x}$ as an initial solution passed to the second phase of the method.

In the second phase, we apply a randomized local search heuristic. Given a permutation $\pi$, represented by a binary matrix ${\bf x}$, we compute the maximum regret $Z({\bf x})$ using program \eqref{mip-1:obj}--\eqref{mip-1:7}. In consecutive iterations, we swap two randomly selected jobs in $\pi$, obtaining a permutation $\pi'$, and compute the corresponding maximum regret $Z(\pi')$. Keeping track of the lowest value of maximum regret encountered so far, we either repeat the procedure by swapping the next pair of randomly selected jobs, if the new value is no higher than the current one, or otherwise we retract to the previous permutation $\pi$, by returning the previously swapped jobs to their previous positions. 

The two-phase procedure can be summarized as follows:
\begin{enumerate}
	\item {\it (phase 1) } Solve the mixed-integer program \eqref{mip-2:obj}--\eqref{mip-2:7}, obtaining fractional $\tilde{{\bf z}}$ and $\tilde{\bf q}$, and binary ${\bf x}_0$.
	\item Repeat for $M$ iterations:
    \begin{enumerate}
		\item Round $z_j = 1$ with probability $\tilde{z}_j$, and $q_j=1$ with probability $\tilde{q_j}$.
		\item For binary ${\bf z}$ and ${\bf q}$ determine ${\bf x}$ feasible for the set of constraints \eqref{mip-1:1}--\eqref{mip-1:7}.
		\item If $Z({\bf x}) < Z({\bf x}_0)$ then put ${\bf x}_0 \leftarrow {\bf x}$.
	\end{enumerate}
	\item {\it (phase 2) } Let $\pi$ be a permutation corresponding to ${\bf x}_0$. Let $S = \{ \pi \}$ and  $\pi^* \leftarrow \pi$.
	\item Repeat for $N$ iterations:
	\begin{enumerate}
		\item Create a schedule $\pi'$ by swapping two randomly selected jobs $i, j$ in $\pi$:
		
		$\pi'(i) = \pi(j)$, $\pi'(j) = \pi(i)$, and $\pi'(k) = \pi(k)$ for all $k \neq i,j$.
		\item If $\pi' \in S$ then discard $\pi'$ and repeat the above step by taking another pair of random $i,j$. Otherwise, $S \leftarrow S \cup \{ \pi' \}$.
		\item If $Z(\pi') < Z(\pi)$ then $\pi^* \leftarrow \pi'$.
		\item If $Z(\pi') \leq Z(\pi)$ then $\pi \leftarrow \pi'$. Otherwise, generate a random real number $r \in [0,1]$. If $r > \alpha$, then $\pi \leftarrow \pi'$.
	\end{enumerate}
	\item Return the schedule $\pi^*$.
\end{enumerate}
The set $S$ is maintained in order to prevent cycling during the search. 
The parameter $\alpha \in [0,1]$ controls the likelihood of proceeding from a worse than previous solution on the search path, and is intended to help avoiding local minima. 
This procedure can be run for prespecified number of iterations $N$, depending on the available computer resources, and can be easily parallelized. Note, however that for large number of jobs, as $N \ll n!$, this methods examines only a very small fraction of the search space.

\section{Experimental Results}

We have examined the solution technique presented in the previous section by comparing it with a simple mid-point heuristic, which is a standard method for tackling min-max regret problems with interval uncertainty \cite{goerigk2016algorithm}. This heuristic outputs a solution of the deterministic counterpart problem with a scenario fixed to interval middle points, $\tilde{p}_i = p_i^- +\frac{1}{2}(p_i^+ - p_i^-)$. Note that for the problem variant with arbitrary weights, this requires solving a knapsack problem.

In each experiment we have generated 10 problem instances for each value of the number of jobs $n$. Each such instance consisted of jobs with processing time intervals generated by taking the lower bound $p_j^-$ as an uniformly random integer between $5$ and $10$, and the upper bound $p_j^+$ by adding to $p_j^-$ and uniformly random integer between $0$ and $20$. Due-dates were uniformly random integers between $5n$ and $10n$. We have considered both unweighted ($w_j=1$ for all $j\in J$) and weighted cases. In the latter, weights are uniformly random integers between $1$ and $100$.

The MIPs used by the solution method were implemented in CPLEX 12.6 software. For larger problem instances the program \eqref{mip-2:obj}--\eqref{mip-2:7} in the phase 1 was usually not solved to optimality; instead, the best feasible solution was returned after running the solver for 60 seconds. However, for all the considered problem instances, program \eqref{mip-1:obj}--\eqref{mip-1:7} was solved to optimality for every fixed permutation. 

For each experiment we report the mean value and the standard deviation of the objective function, estimated from 10 problem instances. Values for both the mid-point scenario heuristic and the proposed method are given. We also report the computation time statistics for our method. Note that these depend on parameters that we have set: $M=100$ in step 2, $N=1000$ and $\alpha = 0.1$ in step 4. 

The results are presented in Tables \ref{tab:1} and \ref{tab:2}. We conclude that the proposed method is consistently better than the mid-point scenario heuristic, especially for the variant of the problem with arbitrary weights.

\begin{table}[h!]
	\caption{Scheduling with the objective to minimize the (unweighted) number of late jobs.}\label{tab:1}
	\centering
\begin{tabular}{|c|cc|ccccc|}
	\hline
     & \multicolumn{2}{c|}{mid-point heuristic} & \multicolumn{5}{c|}{proposed method} \\
$n$	 & mean Z 	 & std Z 	 & mean Z 	& std Z 	& min time 	 & mean time 	 & max time \\
	\hline
10 	 & 2.90		& 0.30 	 	& 2.90 	 & 0.30		 & 73.73 	 & 108.95 	 & 164.79 \\
15 	 & 3.70		& 1.00 		& 3.40 	 & 1.11 	 & 200.48    & 418.10 	 & 618.06 \\
20 	 & 4.70 	& 1.10 		& 4.50 	 & 1.57 	 & 296.33 	 & 408.98 	 & 539.51 \\
25 	 & 5.80		& 1.24 		& 5.20 	 & 2.04 	 & 509.21 	 & 574.94 	 & 632.99 \\
30 	 & 5.70		& 0.90		& 5.60 	 & 1.11 	 & 533.80 	 & 644.97 	 & 679.01 \\
	\hline
\end{tabular}
\end{table}
\begin{table}[h!]
	\caption{Scheduling with the objective to minimize the total weight of late jobs.}\label{tab:2}
	\centering
\begin{tabular}{|c|cc|ccccc|}
	\hline
     & \multicolumn{2}{c|}{mid-point heuristic} & \multicolumn{5}{c|}{proposed method} \\
$n$	 & mean Z 	 & std Z 	 & mean Z 	& std Z 	& min time 	 & mean time 	 & max time \\
	\hline
10 	 & 140.90 	 & 34.65  	& 94.00 	 & 33.74 	 & 66.46 	 & 77.06 	 & 101.70 \\
15 	 & 195.00 	 & 47.04	& 111.80 	 & 50.24 	 & 187.39 	 & 419.93 	 & 886.15 \\
20   & 243.20	 & 67.79 	& 142.56 	 & 72.19 	 & 281.25 	 & 564.59 	 & 637.19 \\
25 	 & 464.50 	 & 118.42   & 155.50 	 & 80.31 	 & 668.37 	 & 728.07 	 & 850.91 \\
30 	 & 444.80 	 & 113.36 	& 149.67 	 & 58.38 	 & 695.58 	 & 914.52 	 & 1132.68 \\
	\hline
\end{tabular}
\end{table}

\section{Conclusions}

Single machine scheduling to minimize the total weight of late jobs with arbitrary processing times and a common due-date is an example of combinatorial problem which is easy to solve if exact values of parameters are known. In practice, this assumption is rarely valid. 
It turns out that interval data min-max regret variant of this problem is much more difficult to solve to optimality. 
We have examined a MIP-based heuristic solution method that successfully handles medium-sized problem instances, and appears to significantly improve on the standard mid-point heuristic. One of the future research directions is the design of efficient approximation methods for the class of problems in question.

\end{document}